\documentclass[authoryear,preprint,review,12pt]{elsarticle}

\usepackage{graphics}

\usepackage{graphicx}

\usepackage{epsfig}

\usepackage{amssymb}

\usepackage{amsthm}

\usepackage{lineno}

\journal{New Astronomy}

\begin{document}

\begin{frontmatter}



\title{A $\gamma$ Doradus Candidate In Eclipsing Binary BD And?}


\author{E. Sipahi\corref{cor1}}
\ead{esin.sipahi@mail.ege.edu.tr}
\cortext[cor1]{Corresponding author}

\author{H. A. Dal}

\address{Ege University, Science Faculty, Department of Astronomy and Space Sciences, 35100 Bornova, \.{I}zmir, Turkey}

\begin{abstract}
The BVR photometric light curves of the eclipsing binary BD And were obtained in 2008 and 2009. We estimated the mass ratio of the system as 0.97 and the photometric solutions were derived. The results show that BD And is a detached binary system, whose components have a little temperature difference of about 40 K. By analyzing photometric available light minimum times, we also derived an update ephemeris and found for the first time a possible periodic oscillation with an amplitude of 0.011 days and a period of 9.6 years. The results indicate that the periodic oscillation could be caused by a third component physically attached to the eclipsing binary. After removing the light variations due to the eclipses and proximity effects, the light-curve distortions are further explained by the pulsation of the primary component with a dominant period of $\sim$1 day. In accordance with the position of the primary component on the Hertzsprung-Russell diagram and its pulsation period, the primary component of BD And could be an excellent $\gamma$ Doradus candidate. It is rarely phenomenon that a component of the eclipsing binary system is a $\gamma$ Doradus variable.
\end{abstract}


\end{frontmatter}



\section{Introduction}

According to the catalogue of parameters for eclipsing binaries \citep{Bra80}, BD And (GSC 03635-01320) is a $\beta$ Lyr type eclipsing binary, which has a spectral type of F8, with a period of $0^{d}.462902$. The binary system BD And was mainly observed to derive minimum times \citep{Kim06, Hub06, Hub07, Hub09a, Hub09b}. \citet{Sha94} included BD And into the catalog of near-contact binary systems, while both \citet{Giu83} and \citet{Mal06} listed the system in their tables of Algol systems. BD And has a blue magnitude $B=11^{m}.6$ as reported in the SIMBAD database and an interesting light curve variation. Its properties are relatively poorly known compared to those of other short-period binaries. There is no spectroscopic study for BD And in the literature.

Although BD And is a well observed system and many minima times have been derived, there is no completed light curve in the literature. Therefore, this binary was included in our observing plan for understanding the properties of the light variation and studying the period change. In this paper, the photometric light curves of BD And in the BVR-bands were observed in 2008 and 2009. Those observations were used to determine the photometric solution. At the same time, orbital period changes were analyzed with all photometric light minimum times. This is the first detailed study of this interesting binary.

\section{Observations}

Observations were acquired with a thermoelectrically cooled ALTA $U+47$ $1024\times1024$ pixel CCD camera attached to a 35 cm - Schmidt - Cassegrains type MEADE telescope at Ege University Observatory. The observations made in BVR bands were continued in August 20, 23, 24, 25, September 04, 15 and October 12, 13, 20 in the season of 2008, while we continued the observations in August 03, 04, 10, 13, 18, 25, 26, 29, 30, September 28 and October 07, 08 in the season of 2009. Some basic parameters of the program stars are listed in Table 1. In the table, GSC 3635 1816 and GSC 3635 838 were observed as comparison stars. Although BD And and the comparison stars are very close in the sky, differential atmospheric extinction corrections were applied. The atmospheric extinction coefficients were obtained from observations of the comparison stars on each night. Heliocentric corrections were also applied to the times of the observations. The mean averages of the standard deviations are $0^{m}.021$, $0^{m}.013$, and $0^{m}.019$ for observations acquired in the BVR bands, respectively. To compute the standard deviations of observations, we used the standard deviations of the reduced differential magnitudes in the sense comparisons minus check stars for each night. 

All the data of BD And were phased by using the minimum time and period taken from \citet{Kre04}. Then, the V-band light curve shown in Figure 1 was derived. The V-band light curve phased with the light-elements given in the literature demonstrated  some clues about the properties of the system, as following: (1) There is only one minimum. However, (2) This minimum seems to be two groups of minima, a shallow one inside a deeper one. (3) There is also light variability at maxima phases.

\section{Light Curve Analysis}

\subsection{A Model with The First Approach}

We took JHK brightness of the system ($J=9^{m}.504$, $H=9^{m}.164$, $K=9^{m}.055$) from the NOMAD Catalogue \citep{Zac04}. Using these brightness, we derived dereddened colours as a $(J-H)_{0}=0^{m}.185$ and $(H-K)_{0}=0^{m}.040$ for the system. Using the calibrations given by \citet{Tok00}, we derived the temperature of the primary component as 7000 K depending on these dereddened colours, which indicate that its spectral type is F2V. Photometric analysis of BD And was carried out using the PHOEBE V.0.31a software \citep{Pra05}, which depends on the method used in the version 2003 of the Wilson-Devinney Code \citep{Wil71, Wil90}.

The BVR light curves, which were phased with the light-elements given by \citet{Kre04}, were analysed simultaneously with the "detached system", "semi-detached system with the primary component filling its Roche-Lobe" and "semi-detached system with the secondary component filling its Roche-Lobe" modes. In the process of the computation, we initially adopted the following fixed parameters: the mean temperature of the primary component ($T_{1}$), the gravity-darkening exponents of $g_{1}=g_{2}=1.0$ \citep{Luc67} and the bolometric albedo coefficients of $A_{1}=A_{2}=1.0$ \citep{Ruc69}. The commonly adjustable parameters employed are the orbital inclination ($i$), the mean temperature of the secondary component ($T_{2}$), the potentials of the components ($\Omega_{1}$ and $\Omega_{2}$) and the monochromatic luminosity of the primary component ($L_{1}$).

The mass ratio $q$ can be estimated using the empirical relation \citep{Kju06};

\begin{center}
\begin{equation}
q~=~\frac{M_{2}}{M_{1}}~\approx~ (\frac{T_{2}}{T_{1}})^{1.7}
\end{equation}
\end{center}
However, the secondary minimum does not appear in the light curves constructed with the light elements given in the literature. Because of this, using different $q$ values from 0.1 to 2, we tried to find an acceptable solution. During the analyses, the first attempts indicated that the secondary component of the system should be a faint-late type star. So, we searched the solution in detail for $q$ values in range of 0.1 and 0.6. On the other hand, we could not reach a unique solution for none of the different $q$ values from 0.1 to 2. None of the obtained solutions can not be statistically acceptable in the astrophysical sense. This result demonstrated that there is a problem in phasing of the data due to the wrong minimum time or especially the wrong period. Considering the absence of the secondary minimum in the light curve (Fig. 1), we decided that the wrong parameter must be the period.

\subsection{The Solution with Real Period}

Considering minima separation into two groups in the primary minima of the light curve phased with the light-elements given in the literature, we examined all consecutive minima one by one. We saw that the depths of minima are regularly changing from one cycle to the next one. It is clear that the orbital period should be two times of those given in the literature. The most remarkable amplitude variation should be seen at the secondary minimum. This is because the pulsating component is the primary one and this pulsating primary component passes in front of the secondary one during the secondary minimum. The effect of the pulsation is not obviously seen around the primary minimum because the non-pulsating secondary component partly covers the pulsating component. Using $0^{d}.925804$ value for the period, all the data were re-phased. Then, we re-derived the light curve, which is shown in the upper panel of Figure 2. The shape of the light curve agrees with the Algol type as was published by \citet{Mal06}.

In this case, as seen in Figure 2, there are two minima with a little amplitude difference in the light curve, one of them lies at the phase of 1.0, while the second one lies at the phase of 1.5. In addition, the low amplitude variation seen at out-of-eclipses in Figure 1 exhibits itself better than the previous. Moreover, the low amplitude is seen around the second minimum from the phase of 1.2 to 1.7. In this light curve, although there is not any level difference among the primary minima, the levels of the secondary minima are still varying from one cycle to the next one. This variation should be a part of the low amplitude variation seen around the secondary minimum, because they all are following themselves with the same low amplitude. The part of this variation in the secondary minimum is shown in the bottom panel of Figure 2. The light variability seen both at the maxima and in the secondary minimum should be originated due to the primary star. 

In this configuration, the preliminary analysis indicates that the system is a detached binary. Therefore, "detached mode" was applied to our analysis. The Wilson-Devinney code is based on Roche geometry which is sensitive to the mass ratio. To find a photometric mass ratio, the solutions were obtained for a series of fixed values of the mass ratio from $q=0.70$ to 1.4 in increments of 0.05. The sum of the squared residuals, ($\Sigma res^{2}$), for the corresponding mass ratios are plotted in Figure 3, where the lowest value of ($\Sigma res^{2}$) was found at $q=0.97$. Therefore, it indicates that the most likely mass ratio appears to be approximately 0.97. The photometric solutions for 0.97 are listed in Table 2 and corresponding light curves are plotted with black lines in Figure 4. The solution parameters obtained from the BVR light curves analyses seem to be acceptable from the astrophysical point of view.

\section{Discussion and Conclusions}

\subsection{Estimated Absolute Parameters and Evalutionary Status}

Although there is not any radial velocity curve of the system, we tried to estimate the absolute parameters of the components. Considering its spectral type, we computed the mass of the primary component, using the calibrations given by \citet{Tok00}, and the mass of the secondary component was calculated from the estimated mass ratio of the system. Using Kepler's third law, we first calculated the semi-major axis ($a=5.79$ $R_{\odot}$), and then the mean radii of the components. All the estimated absolute parameters are listed in Table 3. In Figure 5, we plotted the distribution of the radii versus the masses for the components of the system. In the figure, the open circle represents the secondary component, while the filled circle represents the primary component. The lines represent the ZAMS and TAMS theoretical model developed by \citet{Pol98}.

\subsection{The Light Variations Out-of-Eclipses}

Intrinsic light variations are clearly seen at all the phases without the primary minimum. This indicates that the primary component of the system is responsible for this variation. The light variation of BD And consists of proximity effects, eclipses, and intrinsic variations due from the primary component. The eclipses and the proximity effects in the light curves can be calculated from light curve analysis. The intrinsic light variations less affect these quantities. After extracting the synthetic light curves from the BVR light curves as seen in Figure 4, we looked for frequencies, which could arise from the outside light variation. The frequency analysis was performed using PERIOD04 \citep{Len05}. The peaks were searched for a range from 0 to 167 cd$^{-1}$. No relevant features have been detected at frequencies higher than $\sim$10 cd$^{-1}$, up to the Nyquist limit of $\sim$167 cd$^{-1}$. The prominent features are situated in the frequency interval 0.7-3.2 cd$^{-1}$. In the analysis, we found just one dominant frequency in BVR-bands and these frequencies are listed in Table 4, in which the amplitudes, phases and their corresponding errors are given. The amplitude spectrum of the dominant frequency of BD And data in the BVR-bands are shown in Figure 6. Unfortunately, the BVR data are not adequate for searching the existence of other possible frequencies.

The dominant period of the low amplitude light variations caused by the primary component is about 0$^{d}$.9988. This case indicates that the primary component is a candidate for $\gamma$ Dor type pulsating stars. According to Kaye et al. (1999), $\gamma$ Dor type pulsation causes a sinusoidal light curves with amplitudes of a few $mmag$ to a few per cent in a time-scales of 0$^{d}$.4 - 3$^{d}$.0. These type stars generally exhibit multi-periodic pulsations. Although the $\gamma$ Doradus stars are located on the main-sequence together with $\delta$ Scuti star, they are generally located especially around the cool border of $\delta$ Scuti star instability strip. They also have a different pulsation mechanism compared to $\delta$ Scuti stars \citep{Han02}. According to \citet{Hen05}, there are three points to identify a star as a $\gamma$ Doradus variable: (1) the stars spectral type should be a late A or early F, (2) luminosity class should be IV or V, (3) the star should have periodic photometric variability in the $\gamma$ Doradus period range that is attributable to pulsation.

Our results obtained in this study demonstrate that the primary component exhibits all of them. The relationship between the $\delta$ Scuti and $\gamma$ Doradus pulsators is a debating subject \citep{Han02}. Handler and Shobbrook (2002) indicated that the pulsation constants ($Q$) is a parameter to classify a star as $\delta$ Scuti or $\gamma$ Doradus stars. Using the equation (2) taken from Handler and  Shobbrook (2002), we found a pulsation constant for the primary component of BD And as 0$^{d}$.74. The case also indicates that the primary component is a $\gamma$ Doradus type pulsator. The primary component of BD And is plotted as an asteriks together with known $\gamma$ Doradus stars, whose data were taken from \citet{Hen05}, in Figure 7. The pulsating primary component of the BD And lies just at the middle of the instability strip of $\gamma$ Doradus type pulsators. Our results reveal that the primary component of BD And could be a $\gamma$ Doradus type variable. Its physical characteristics are the same with the nature of VZ CVn \citep{Iba07}. VZ CVn is the only known system of this type.

\begin{center}
\begin{equation}
\log Q_{\rm i}~=~C~+~0.5\log g~+~0.1M_{\rm bol}~+~\log T_{\rm eff}~+~\log P_{\rm i}
\end{equation}
\end{center}

\subsection{The O-C Variation}

In the literature, 116 minima times in total have been obtained for the system. 73 minima times were obtained from photoelectric and CCD observations, while the rest are photographic and visual measurements. The minima times obtained photographically have very large standard deviation. Using the SPSS V17.0 software \citep{Gre99} and GrahpPad Prism V5.02 software \citep{Daw04, Mot07}, we statistically analysed all the minima times to search whether there is any regular variation, or not. In the first step, we analysed all the data together, but the obtained correlation coefficients of the fits were found to be very low. In addition, $p-values$ were found to be higher than the threshold level of significance ($\alpha-value$) of 0.005. As it is well known that both $p-~and~\alpha-values$ are used to test whether the obtained fit is statistically acceptable, or not \citep{Wal03}. In the second step, we separated the data into the groups, as photoelectric, CCD, visual and photographical observations. Then, both groups of the data were analysed separately. According to $p-value$, although an acceptable fit was obtained just from the photoelectric and CCD observations, no acceptable fit was found from the photographic and visual observations. As a results, we used just photoelectric and CCD observations in the analyses. 

All minima were taken from the literature and from the more recent studies listed in Table 5. We obtained fourteen new epochs of minimum during the observations and seven of them have already been published by \citet{Sip09}. We used these minima times listed in Table 5 to determine the light elements by using the least squares method. We have found the new light elements as following:

\begin{center}
\begin{equation}
JD~(Hel.)~=~24~55057.4656(5)~+~0^{d}.9258051(1)~\times~E.
\end{equation}
\end{center}
We plot the residuals of all the eclipse timings as $(O-C)$ in the upper panel of Figure 8. 

There is a sinusoidal oscillation in the O-C residuals for BD And. The oscillating characteristic may be caused by the light-time effect due to the existence of the third body orbiting the eclipsing binary or a possible result of magnetic activity cycles of the system. Since no spectroscopic solutions are available in the literature of BD And, its absolute parameters can not be directly determined. We estimated the primary mass as 1.56 $M_{\odot}$, and the radius as 1.42 $R_{\odot}$ by assuming the primary component to be a normal and main-sequence F2 star. Based on our photometric solutions, the mass of the secondary component was found to be $M_{2}=1.51$ $M_{\odot}$, while the separation ($a$) between the two components was calculated as 5.79 $R_{\odot}$. Both the observed colours and the light curve solution of the system show that the components are not late type stars. Therefore, the magnetic activity cycle is not a possible mechanism to cause the period variation. However, the BVR light curve analysis gives a physically acceptable values for a third light contribution (see Table 2). Therefore, the period oscillation may be caused due to the light-time effect. A weighted least-squares solution for two parameters ($T_{0}$ and P) of the linear ephemeris of BD And and five parameters ($a_{12}sini'$, $e'$, $w'$, $T'$ and $P'$) of the light-time effect are presented in Table 6. The observational points and theoretical best fit curve are plotted against epoch number in Figure 8. The parameters we derived for the light-time effect enable calculations of the mass function, using the Equation (4) taken from \citet{Kop78}:

\begin{center}
\begin{equation}
f(m)~=~\frac{(a_{12}sini')^{3}}{(P')^{2}}=\frac{(M_{3}~sini')^{3}}{(M_{1}+ M_{2}+ M_{3})^{2}}
\end{equation}
\end{center}
where $M_{1}$, $M_{2}$ and $M_{3}$ are the masses of the binary, and the third body, respectively. We obtained the mass function as $f(m)=0.0089$ $M_{\odot}$ for the third body. The mass of the third component can be estimated with Equation (4), depending on the orbital inclination. We calculated the third body masses at different inclination, $i'$ values, which are shown in Table 6. Here, the total mass of the eclipsing system was taken as 3.07 $M_{\odot}$. The result parameters given in Table 6 suggest that BD And has an eccentric orbit around the mass center of the third-body system with a period of $\sim$9.6 years.

Furthermore, a significant contribution of the third light was found in the light curve analysis. Thus, it confirms that third body may produce the sinusoidal variation of the period. It is more likely that this third light, used in our light curve solutions, was caused by a third body in the system. Although BD And was included into the list of near-contact binary by \citet{Sha94}, our results showed that BD And is a detached binary system with a little temperature difference of about 40 K between the two components. BD And seems to be an analogue of those detached systems with a possible third body, such as V2080 Cyg \citep{Iba08}.

The aim of the present paper is to draw attention to eclipsing binary BD And which shows interesting features photometrically. Of course, our solutions are based on photometric observations only. For better understanding of the properties and the evolutionary state of BD And, spectroscopic observations are needed.

\section*{Acknowledgments} The authors wish to thank Prof. Dr. \"{O}mer L\"{u}tfi De\v{g}irmenci for his help with the $O-C$ analysis. The authors acknowledge generous allotments of observing time at the Ege University Observatory. We also thank the referee for useful comments that have contributed to the improvement of the paper.

\clearpage

\begin{figure}
\hspace{1cm}
\includegraphics[width=15.5cm]{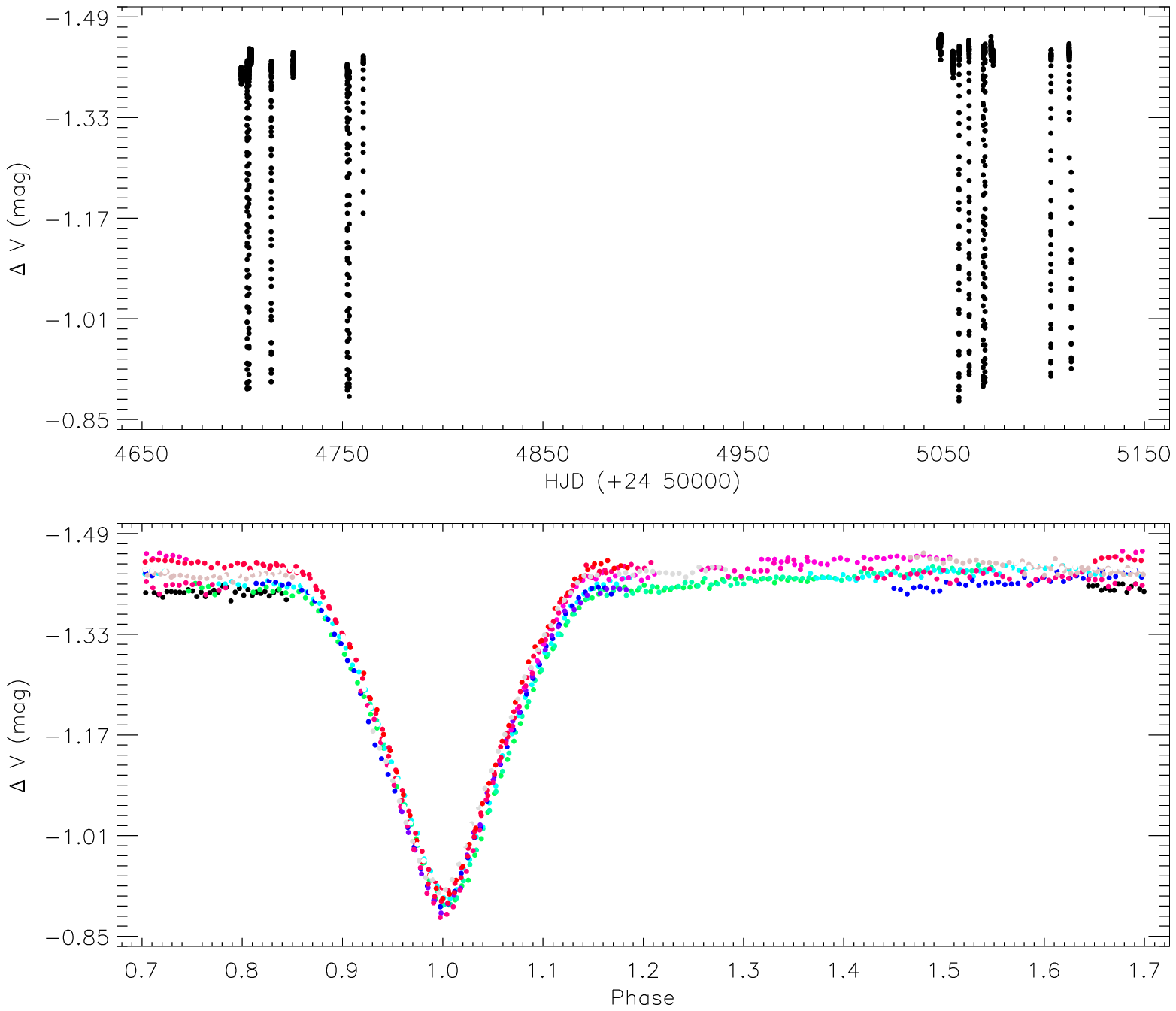}
\vspace{0.02cm}
\caption{Upper and lower panels show the differential magnitudes versus time (in HJD) and phase. Here, the data were phased by using the light-elements given by \citet{Kre04}.}
\label{fig:Fig.1}
\end{figure}

\begin{figure}
\hspace{1.8cm}
\includegraphics[width=22cm]{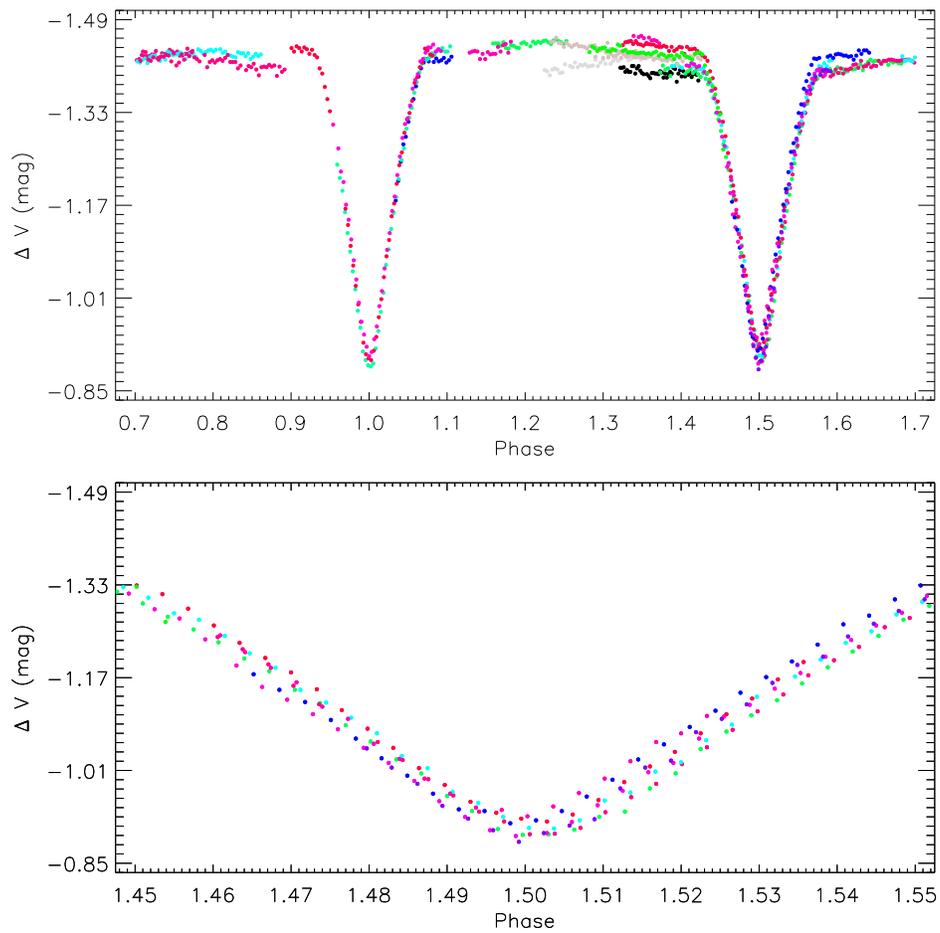}
\vspace{0.02cm}
\caption{The V-band light curve is shown in the upper panel. Here, the data were phased by using twice the period given by \citet{Kre04}. In the bottom panel, the secondary minima are plotted for a closer look to the minima for better visibility of light variations. In the figure, the colours represent observations obtained in different nights.}
\label{fig:Fig.2}
\end{figure}

\begin{figure}
\hspace{1.6cm}
\includegraphics[width=16cm]{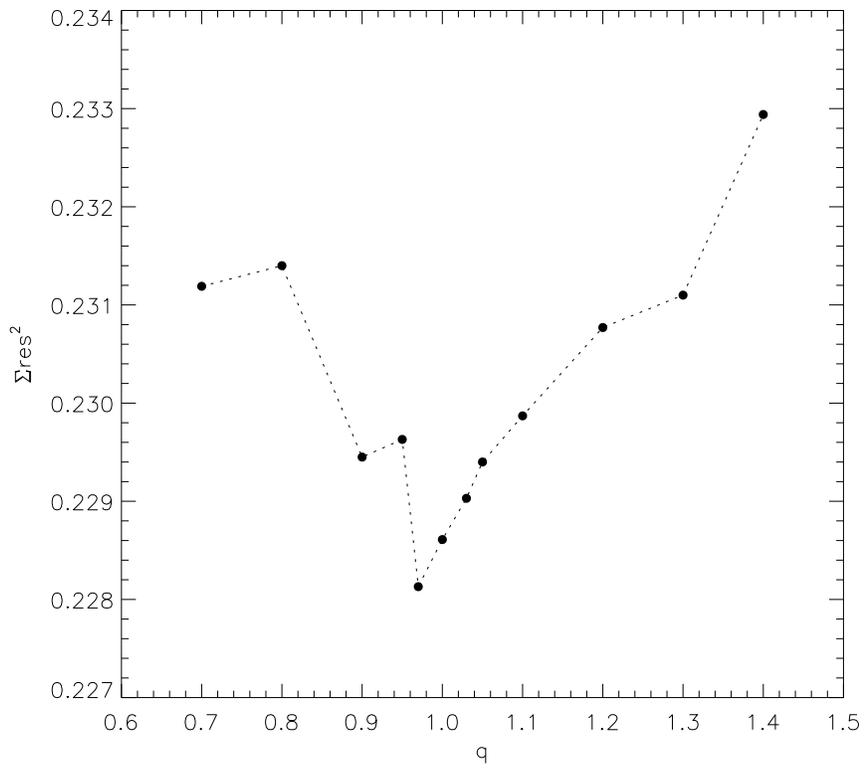}
\vspace{0.02cm}
\caption{The variation of the sum of weighted squared residuals versus mass ratio in the "q search".}
\label{fig:Fig.3}
\end{figure}

\begin{figure}
\includegraphics[width=15cm]{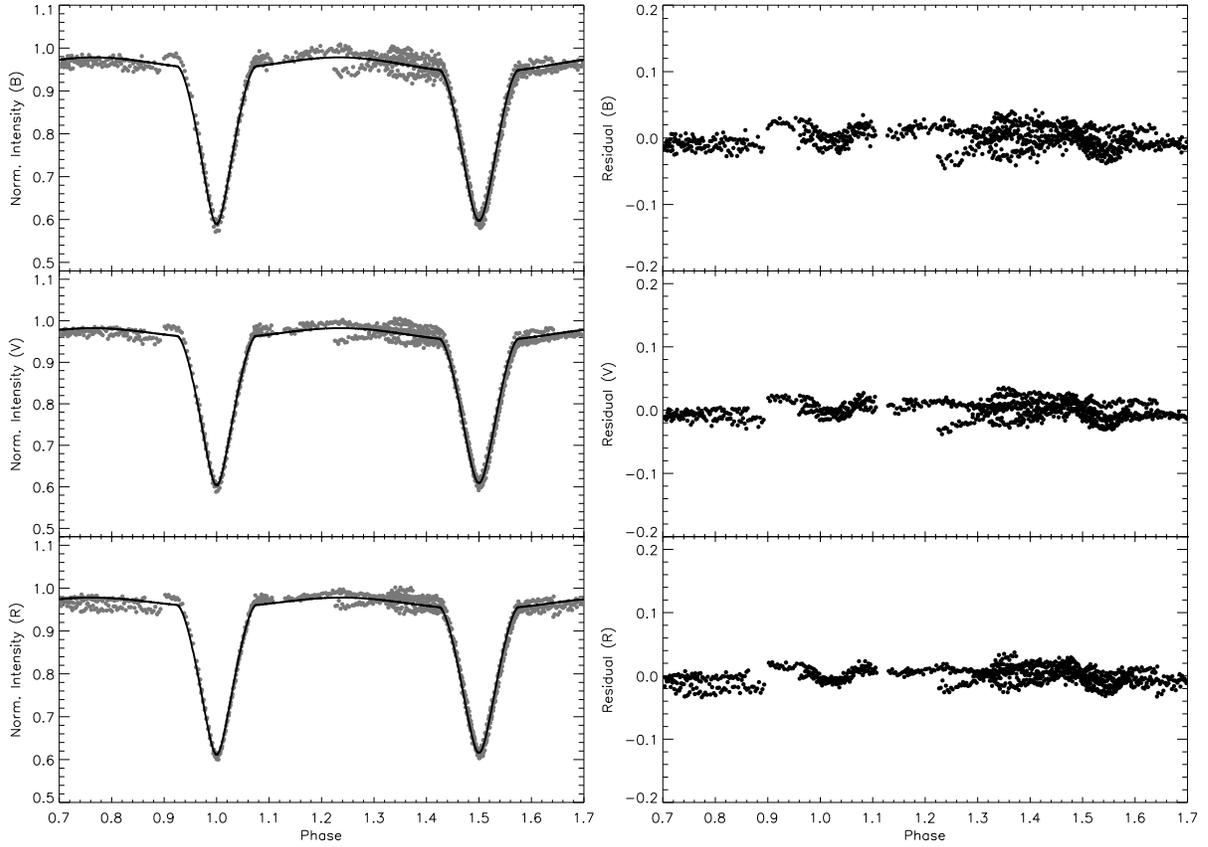}
\vspace{0.02cm}
\caption{BD And's light curves (dimmer - filled circle) observed in BVR bands and the synthetic curves (line) derived from the light curve solutions in each band are shown in the left side panels. The prewhitening curves (dark filled circle) obtained from the residual data are shown in the right side panels.}
\label{fig:Fig.4}
\end{figure}

\begin{figure}
\hspace{2.85cm}
\includegraphics[width=20cm]{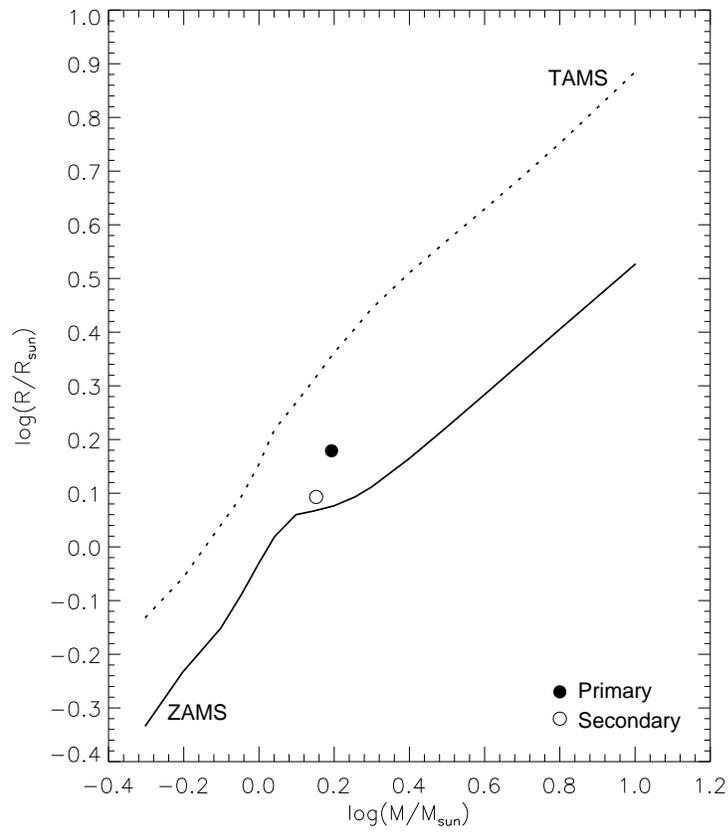}
\vspace{0.02cm}
\caption{The places of the components of BD And in the Mass-Radius distribution. The black-line and dashed line represent the ZAMS and TAMS theoretical models developed by Pols et al. (1998).}
\label{fig:Fig.5}
\end{figure}

\begin{figure}
\hspace{1.65cm}
\includegraphics[width=28cm]{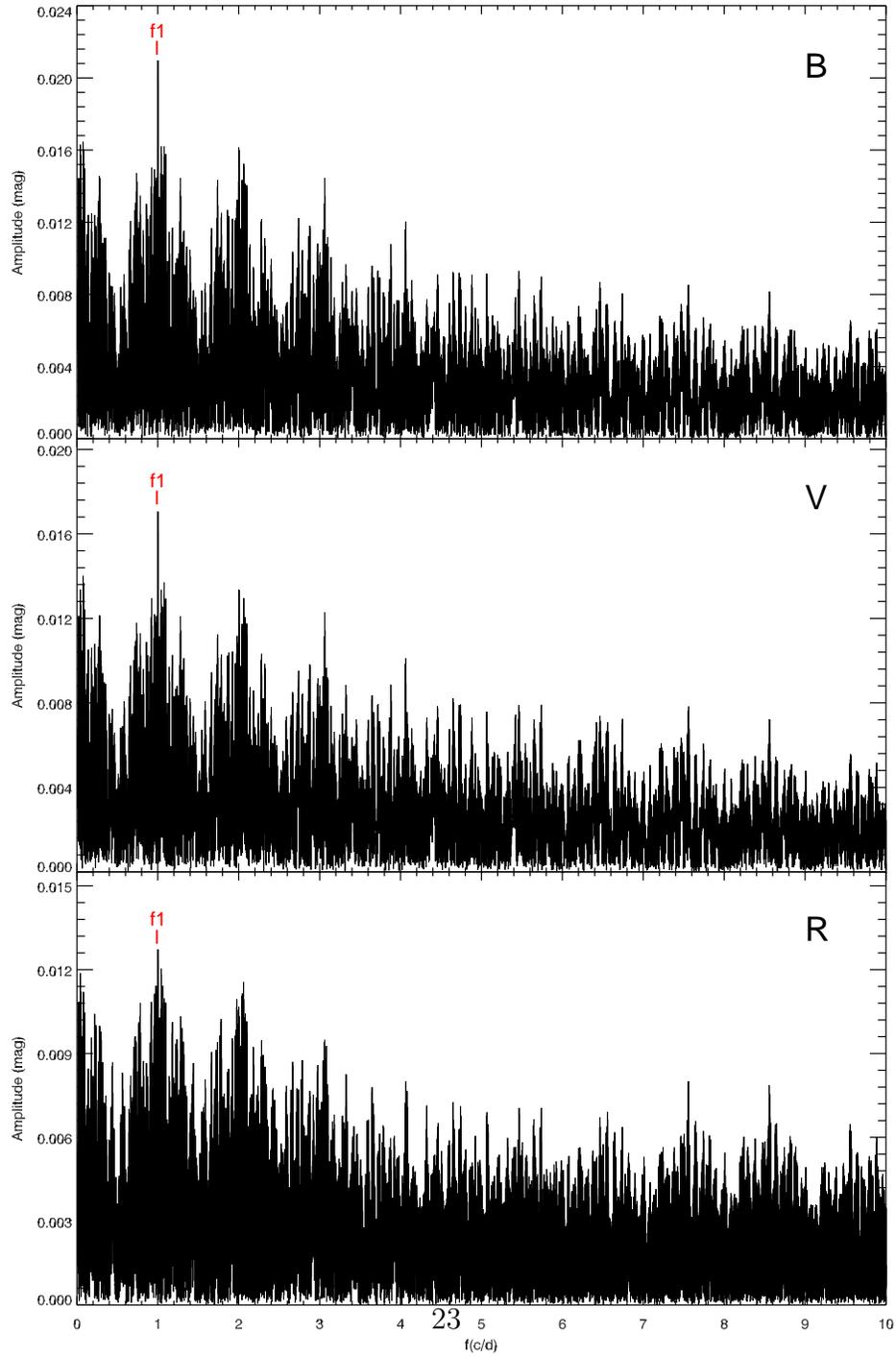}
\vspace{0.02cm}
\caption{The result of the frequency analysis are shown from the upper to the bottom for the pulsating component of BD And in the BVR filter, respectively. High frequency spectrum of the residuals were obtained after removing the binary model.}
\label{fig:Fig.6}
\end{figure}

\begin{figure}
\hspace{1.7cm}
\includegraphics[width=15cm]{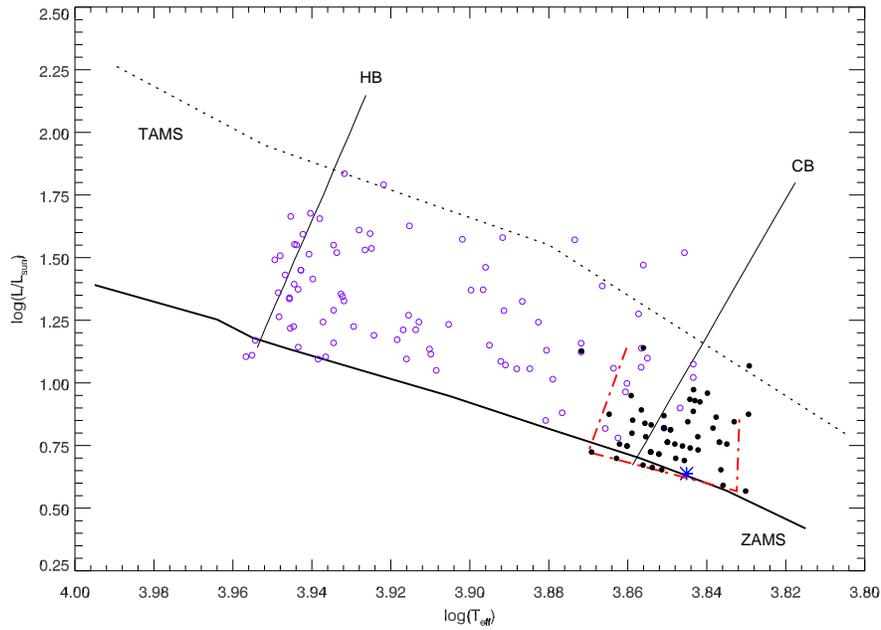}
\vspace{0.02cm}
\caption{The place of the primary component of BD And among $\gamma$ Doradus type stars in the HR diagram. In the figure, the small filled circles represent $\gamma$ Doradus type stars listed in \citet{Hen05}. The asterisk represents the primary component of the system. The dash dotted lines (red) represent the borders of the area, in which $\gamma$ Doradus type stars take place. In addition we plotted the hot (HB) and cold (CB) borders of the $\delta$ Scuti stars for comparison. In the figure, the purpel open circles represent some semi- and un-detached binaries taken from \citet{Soy06} and references therein. The ZAMS and TAMS were taken from \citet{Gir00}, while the borders of the Instability Strip were computed from \citet{Rol02}.}
\label{fig:Fig.7}
\end{figure}

\begin{figure}
\hspace{1.2cm}
\includegraphics[width=16cm]{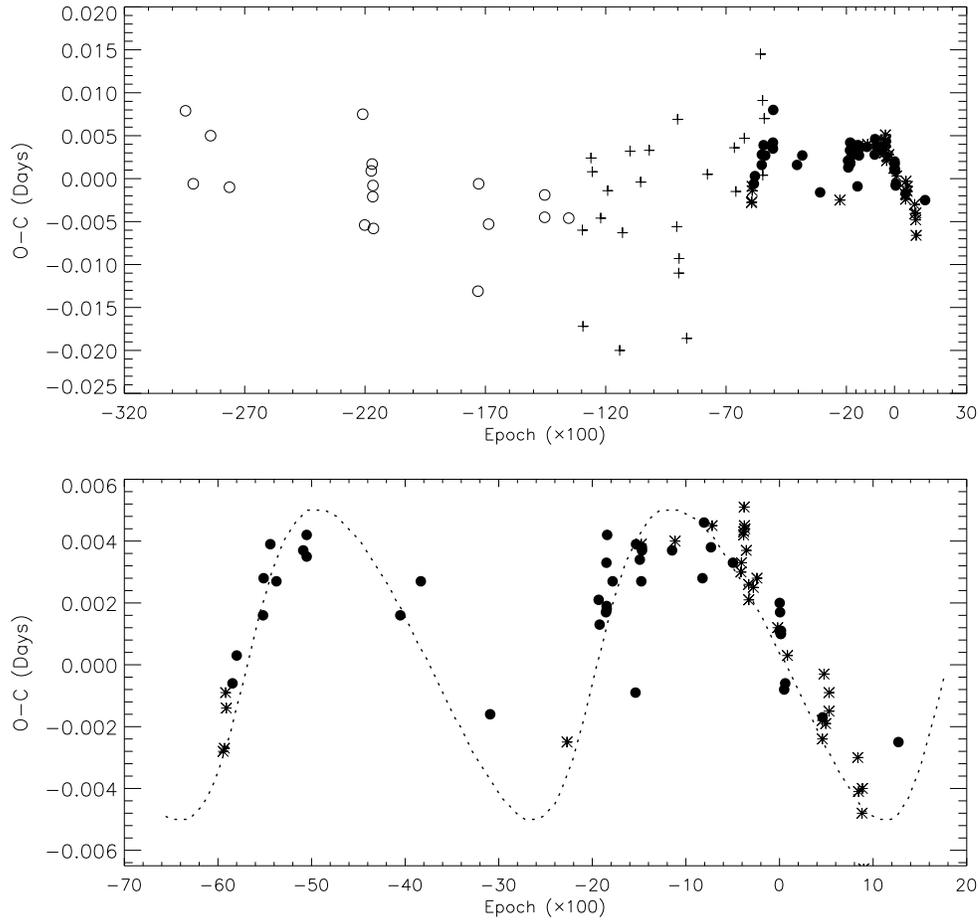}
\vspace{0.02cm}
\caption{The O-C diagram depending only on observations of BD And. In the upper panel, all available O-C data are shown, while the O-C data used in the analyses are shown in the bottom panel. In both panels, the filled circles represent the CCD observations, while the asterisk represents the photoelectric observations. In bottom panel, the dashed curve represents the predicted light-time effect. The open circles represent the photographic observations, while the plus represent the visual observations.}
\label{fig:Fig.8}
\end{figure}

\clearpage

\begin{table*}
\centering
\caption{Basic parameters for the observed stars.}
\begin{tabular}{lccccc}
\hline\hline
Star & Alpha (J2000) & Delta (J2000) & V \\
& ($^{h}$ $^{m}$ $^{s}$) & ($^{\circ}$ $^{\prime}$ $^{\prime\prime}$) & (mag) \\
\hline
BD And & 23 07 05.162 & +50 57 30.95 & 10.84 \\
GSC 3635 1816 & 23 07 09.720 & +50 57 35.80 & 12.20 \\
GSC 3635 838 & 23 07 11.290 & +50 55 23.40 & 11.70 \\
\hline
\end{tabular}
\end{table*}

\begin{table*}
\centering
\caption{The parameters obtained from the light curve analysis.}
\begin{tabular}{lr}
\hline\hline
Parameter & Value \\
\hline
$q$ & 0.97 \\
$i$ ($^\circ$) & 85.88$\pm$0.07 \\
$T_{1}$ (K) & 7000 \\
$T_{2}$ (K) & 6962$\pm$10 \\
$\Omega_{1}$ & 5.108$\pm$0.012 \\
$\Omega_{2}$ & 5.585$\pm$0.019 \\
$L_{1}/L_{T}(B)$ & 0.570$\pm$0.048 \\
$L_{1}/L_{T}(V)$ & 0.567$\pm$0.046 \\
$L_{1}/L_{T}(R)$ & 0.565$\pm$0.045 \\
$L_{2}/L_{T}(B)$ & 0.422$\pm$0.048 \\
$L_{2}/L_{T}(V)$ & 0.424$\pm$0.046 \\
$L_{2}/L_{T}(R)$ & 0.424$\pm$0.045 \\
$L_{3}/L_{T}(B)$ & 0.008$\pm$0.037 \\
$L_{3}/L_{T}(V)$ & 0.009$\pm$0.038 \\
$L_{3}/L_{T}(R)$ & 0.011$\pm$0.039 \\
$g_{1}$, $g_{2}$ & 1.0, 1.0 \\
$A_{1}$, $A_{2}$ & 1.0, 1.0 \\
$x_{1,bol}$, $x_{2,bol}$ & 0.471, 0.473 \\
$x_{1,B}$, $x_{2,B}$ & 0.599, 0.609 \\
$x_{1,V}$, $x_{2,V}$ & 0.495, 0.501 \\
$x_{1,R}$, $x_{2,R}$ & 0.399, 0.404 \\
$<r_{1}>$ & 0.246$\pm$0.001 \\
$<r_{2}>$ & 0.215$\pm$0.001 \\
\hline
\end{tabular}
\end{table*}

\begin{table*}
\centering
\caption{The estimated absolute parameters derived for BD And system.}
\begin{tabular}{lcc}
\hline\hline
Parameter & Primary & Secondary \\
\hline
Mass ($M_{\odot}$) & 1.56 & 1.51 \\
Radius ($R_{\odot}$) & 1.42 & 1.24 \\
Luminosity ($L_{\odot}$) & 4.34 & 3.24 \\
$M_{bol}$ (mag) & 3.15 & 3.46 \\
$log~g$ & 4.33 & 4.43 \\
\hline
\end{tabular}
\end{table*}

\begin{table*}
\centering
\caption{The results of the frequency analysis.}
\begin{tabular}{cccccc}
\hline\hline
Filter	& &	Frequency (c/d)	&	Amplitude (mag)	&	Phase	& SNR \\
\hline
B & $f_{1}$	&	1.00138$\pm$0.00007	&	0.0210$\pm$0.0006	&	0.63$\pm$0.01	& 9.03	\\
V & $f_{1}$	&	1.00125$\pm$0.00008	&	0.0171$\pm$0.0003	&	0.25$\pm$0.01	& 9.16	\\
R & $f_{1}$	&	1.00129$\pm$0.00007	&	0.0127$\pm$0.0006	&	0.03$\pm$0.02	& 8.32	\\
\hline
\end{tabular}
\end{table*}

\begin{table*}
\centering
\caption{Times of minimum light for BD And.}
\begin{tabular}{ccccl}
\hline\hline
HJD (+24 00000) & Cycle & $(O-C)$ & Method & Ref. \\
\hline
27784.1790	&	-29459.0	&	0.0142	&	PG	&	\citet{Kre04}	\\
28081.3540	&	-29138.0	&	0.0056	&	PG	&	\citet{Kre04}	\\
28749.3280	&	-28416.5	&	0.0110	&	PG	&	\citet{Kre04}	\\
29476.0790	&	-27631.5	&	0.0048	&	PG	&	\citet{Kre04}	\\
34600.4190	&	-22096.5	&	0.0119	&	PG	&	\citet{Aza56} 	\\
34677.2480	&	-22013.5	&	-0.0009	&	PG	&	\citet{Aza56} 	\\
34930.4620	&	-21740.0	&	0.0053	&	PG	&	\citet{Aza56} 	\\
34962.4030	&	-21705.5	&	0.0060	&	PG	&	\citet{Aza56} 	\\
34987.3960	&	-21678.5	&	0.0023	&	PG	&	\citet{Aza56} 	\\
34993.4150	&	-21672.0	&	0.0035	&	PG	&	\citet{Aza56} 	\\
35012.3890	&	-21651.5	&	-0.0015	&	PG	&	\citet{Aza56} 	\\
39035.4680	&	-17306.0	&	-0.0098	&	PG	&	\citet{Kre04}	\\
39061.4030	&	-17278.0	&	0.0026	&	PG	&	\citet{Kre04}	\\
39443.2930	&	-16865.5	&	-0.0021	&	PG	&	\citet{Kre04}	\\
41599.4940	&	-14536.5	&	-0.0019	&	PG	&	\citet{Kre04}	\\
41602.2740	&	-14533.5	&	0.0007	&	PG	&	\citet{Kre04}	\\
42525.2990	&	-13536.5	&	-0.0023	&	PG	&	\citet{Kre04}	\\
43050.6920	&	-12969.0	&	-0.0039	&	VIS	&	\citet{Bal96}	\\
43073.8260	&	-12944.0	&	-0.0150	&	VIS	&	\citet{Bal96}	\\
43380.7500	&	-12612.5	&	0.0045	&	VIS	&	\citet{Bal96}	\\
43429.8160	&	-12559.5	&	0.0028	&	VIS	&	\citet{Bal96}	\\
43755.6940	&	-12207.5	&	-0.0027	&	VIS	&	\citet{Bal96}	\\
44022.7920	&	-11919.0	&	0.0005	&	VIS	&	\citet{Bal96}	\\
44485.6760	&	-11419.0	&	-0.0182	&	VIS	&	\citet{Bal96}	\\
\hline
\end{tabular}
\end{table*}

\setcounter{table}{4}
\begin{table*}
\centering
\caption{Continued.}
\begin{tabular}{ccccl}
\hline\hline
HJD (+24 00000) & Cycle & $(O-C)$ & Method & Ref. \\
\hline
44593.5460	&	-11302.5	&	-0.0046	&	VIS	&	\citet{Bal96}	\\
44885.6470	&	-10987.0	&	0.0048	&	VIS	&	\citet{Bal96}	\\
45291.6090	&	-10548.5	&	0.0012	&	VIS	&	\citet{Bal96}	\\
45622.5880	&	-10191.0	&	0.0047	&	VIS	&	\citet{Bal96}	\\
46682.6260	&	-9046.0	&	-0.0045	&	VIS	&	\citet{Bal96}	\\
46713.6530	&	-9012.5	&	0.0081	&	VIS	&	\citet{Bal96}	\\
46756.6850	&	-8966.0	&	-0.0099	&	VIS	&	\citet{Bal96}	\\
46769.6480	&	-8952.0	&	-0.0082	&	VIS	&	\citet{Bal96}	\\
47062.6560	&	-8635.5	&	-0.0176	&	VIS	&	\citet{Bal96}	\\
47861.6450	&	-7772.5	&	0.0014	&	VIS	&	\citet{Bal96}	\\
48897.6240	&	-6653.5	&	0.0041	&	VIS	&	\citet{Bal96}	\\
48954.5560	&	-6592.0	&	-0.0009	&	VIS	&	\citet{Bal96}	\\
49278.5940	&	-6242.0	&	0.0052	&	VIS	&	\citet{Bal96}	\\
49554.4764	&	-5944.0	&	-0.0024	&	PE	&	\citet{Age95}	\\
49567.4378	&	-5930.0	&	-0.0023	&	PE	&	\citet{Age95}	\\
49578.5492	&	-5918.0	&	-0.0005	&	PE	&	\citet{Age95}	\\
49585.4923	&	-5910.5	&	-0.0010	&	PE	&	\citet{Age95}	\\
49646.5962	&	-5844.5	&	-0.0002	&	CCD	&	\citet{Hub06}	\\
49688.2583	&	-5799.5	&	0.0006	&	CCD	&	\citet{Hub06}	\\
49899.8190	&	-5571.0	&	0.0148	&	VIS	&	\citet{Bal96}	\\
49948.4109	&	-5518.5	&	0.0019	&	CCD	&	\citet{Hub06}	\\
49954.4298	&	-5512.0	&	0.0031	&	CCD	&	\citet{Hub06}	\\
49983.5990	&	-5480.5	&	0.0094	&	VIS	&	\citet{Bal96}	\\
49989.6080	&	-5474.0	&	0.0007	&	VIS	&	\citet{Bal96}	\\
\hline 
\end{tabular}
\end{table*}

\setcounter{table}{4}
\begin{table*}
\centering
\caption{Continued.}
\begin{tabular}{ccccl}
\hline\hline
HJD (+24 00000) & Cycle & $(O-C)$ & Method & Ref. \\
\hline
50020.6260	&	-5440.5	&	0.0042	&	CCD	&	\citet{Bal96}	\\
50044.7000	&	-5414.5	&	0.0072	&	VIS	&	\citet{Bal96}	\\
50081.2650	&	-5375.0	&	0.0029	&	CCD	&	\citet{Hub06}	\\
50346.5092	&	-5088.5	&	0.0039	&	CCD	&	\citet{Hub06}	\\
50380.3009	&	-5052.0	&	0.0037	&	CCD	&	\citet{Bla96}	\\
50380.3016	&	-5052.0	&	0.0044	&	CCD	&	\citet{Hub06}	\\
50391.4150	&	-5040.0	&	0.0081	&	CCD	&	\citet{Ace97}	\\
51306.5670	&	-4051.5	&	0.0015	&	CCD	&	\citet{Kre04}	\\
51509.3195	&	-3832.5	&	0.0026	&	CCD	&	\citet{Bla00}	\\
52195.3368	&	-3091.5	&	-0.0019	&	CCD	&	\citet{Bla01}	\\
52955.4219	&	-2270.5	&	-0.0030	&	PE	&	\citet{Hub06}	\\
53268.3486	&	-1932.5	&	0.0014	&	CCD	&	\citet{Hub06}	\\
53277.6059	&	-1922.5	&	0.0007	&	CCD	&	\citet{Bal07}	\\
53340.5611	&	-1854.5	&	0.0011	&	CCD	&	\citet{Bal07}	\\
53345.1917	&	-1849.5	&	0.0027	&	CCD	&	\citet{Chu06} 	\\
53347.0419	&	-1847.5	&	0.0013	&	CCD	&	\citet{Chu06} 	\\
53347.9676	&	-1846.5	&	0.0012	&	CCD	&	\citet{Chu06} 	\\
53352.1361	&	-1842.0	&	0.0035	&	CCD	&	\citet{Chu06} 	\\
53406.2942	&	-1783.5	&	0.0020	&	CCD	&	\citet{Hub06}	\\
53632.6500	&	-1539.0	&	-0.0016	&	CCD	&	\citet{Bal07}	\\
53638.2096	&	-1533.0	&	0.0032	&	CCD	&	\citet{Chu06} 	\\
53675.2413	&	-1493.0	&	0.0027	&	CCD	&	\citet{Chu06} 	\\
53690.0535	&	-1477.0	&	0.0020	&	CCD	&	\citet{Chu06} 	\\
53690.0547	&	-1477.0	&	0.0032	&	PE	&	\citet{Kre04}	\\
\hline 
\end{tabular}
\end{table*}

\setcounter{table}{4}
\begin{table*}
\centering
\caption{Continued.}
\begin{tabular}{ccccl}
\hline\hline
HJD (+24 00000) & Cycle & $(O-C)$ & Method & Ref. \\
\hline
53696.0723	&	-1470.5	&	0.0030	&	CCD	&	\citet{Chu06} 	\\
53696.9980	&	-1469.5	&	0.0029	&	CCD	&	\citet{Chu06} 	\\
53994.6444	&	-1148.0	&	0.0029	&	CCD	&	\citet{Bal07}	\\
54024.2705	&	-1116.0	&	0.0032	&	PE	&	\citet{Hub07}	\\
54295.5302	&	-823.0	&	0.0019	&	CCD	&	\citet{Hub09a}	\\
54310.8077	&	-806.5	&	0.0037	&	CCD	&	\citet{Bal07}	\\
54379.3165	&	-732.5	&	0.0029	&	CCD	&	\citet{Hub09b} 	\\
54390.4269	&	-720.5	&	0.0036	&	PE	&	\citet{Hub07}	\\
54596.8802	&	-497.5	&	0.0023	&	CCD	&	\citet{Kre04}	\\
54676.4992	&	-411.5	&	0.0020	&	VIS	&	\citet{Hub10a}	\\
54680.6656	&	-407.0	&	0.0023	&	CCD	&	\citet{Kre04}	\\
54702.4229	&	-383.5	&	0.0032	&	CCD	&	\citet{Sip09}	\\
54703.3488	&	-382.5	&	0.0033	&	CCD	&	\citet{Sip09}	\\
54707.5158	&	-378.0	&	0.0041	&	CCD	&	\citet{Sip09}	\\
54709.3668	&	-376.0	&	0.0035	&	CCD	&	\citet{Sip09}	\\
54714.4586	&	-370.5	&	0.0034	&	CCD	&	\citet{Sip09}	\\
54728.3450	&	-355.5	&	0.0027	&	VIS	&	\citet{Hub10a}	\\
54752.4148	&	-329.5	&	0.0016	&	CCD	&	\citet{Sip09}	\\
54753.3401	&	-328.5	&	0.0011	&	CCD	&	\citet{Sip09}	\\
54798.2421	&	-280.0	&	0.0015	&	PE	&	\citet{Hub09b}	\\
54834.3488	&	-241.0	&	0.0018	&	PE	&	\citet{Hub10a}	\\
55039.4130	&	-19.5	&	0.0001	&	PE	&	\citet{Hub10b}	\\
55057.4661	&	0.0	&	0.0000	&	CCD	&	This Study	\\
55058.8557	&	1.5	&	0.0009	&	CCD	&	\citet{Kre04}	\\
\hline 
\end{tabular}
\end{table*}

\setcounter{table}{4}
\begin{table*}
\centering
\caption{Continued.}
\begin{tabular}{ccccl}
\hline\hline
HJD (+24 00000) & Cycle & $(O-C)$ & Method & Ref. \\
\hline
55062.5586	&	5.5	&	0.0006	&	CCD	&	This Study	\\
55069.5015	&	13.0	&	-0.0001	&	PE	&	This Study	\\
55070.4273	&	14.0	&	-0.0001	&	PE	&	This Study	\\
55071.3532	&	15.0	&	0.0000	&	CCD	&	\citet{Hub10b}	\\
55103.2916	&	49.5	&	-0.0019	&	CCD	&	This Study	\\
55113.4756	&	60.5	&	-0.0017	&	CCD	&	This Study	\\
55135.6958	&	84.5	&	-0.0009	&	PE	&	\citet{Die10} 	\\
55481.4814	&	458.0	&	-0.0036	&	PE	&	\citet{Hub11}	\\
55483.3337	&	460.0	&	-0.0029	&	CCD	&	\citet{Hub11}	\\
55497.6850	&	475.5	&	-0.0016	&	PE	&	\citet{Die11}	\\
55514.3479	&	493.5	&	-0.0032	&	PE	&	\citet{Pas06}	\\
55547.2144	&	529.0	&	-0.0028	&	PE	&	\citet{Pas06}	\\
55547.2150	&	529.0	&	-0.0022	&	PE	&	\citet{Pas06}	\\
55480.5562	&	457.0	&	-0.0030	&	PE	&	\citet{Hub12a}	\\
55832.3609	&	837.0	&	-0.0043	&	PE	&	\citet{Hub12b}	\\
55839.3033	&	844.5	&	-0.0055	&	PE	&	\citet{Hub12b}	\\
55873.5574	&	881.5	&	-0.0062	&	PE	&	\citet{Hub12b}	\\
55877.2615	&	885.5	&	-0.0053	&	PE	&	\citet{Hub12b}	\\
55890.2201	&	899.5	&	-0.0080	&	PE	&	\citet{Die13}	\\
56233.6979	&	1270.5	&	-0.0040	&	PE	&	This Study	\\
\hline 
\end{tabular}
\begin{list}{}{}															
\item[$Note:$]{\small In the fourth column, PG refers to photographic observation, VIS refers to visual observations, PE refers to photoelectric observation, CCD refers to CCD observation.}
\end{list} 
\end{table*}

\begin{table*}
\centering
\caption{The parameters derived from $O-C$ analysis of BD And.}
\begin{tabular}{lr}
\hline\hline
Parameter & Value \\
\hline
$T_{o}$ & 55057.4656(5) \\
$P$ (day) & 0.9258051(1) \\
$T'$ & 53300(90) \\
$P'$ (year) & 9.56(2) \\
$e'$ & 0.35(8) \\
$w'$ ($^\circ$) & 2.76(8) \\
$a_{12}sini'$ (AU) & 0.93(9) \\
$A$ (day) & 0.0054(2) \\
$f(m)$ & 0.0089(6) $M_{\odot}$ \\
$M_{3}$($i'$=$30^\circ$) & 0.99 $M_{\odot}$ \\
$M_{3}$($i'$=$45^\circ$) & 0.70 $M_{\odot}$ \\
$M_{3}$($i'$=$60^\circ$) & 0.58 $M_{\odot}$ \\
$M_{3}$($i'$=$75^\circ$) & 0.50 $M_{\odot}$ \\
$M_{3}$($i'$=$90^\circ$) & 0.47 $M_{\odot}$ \\
\hline 
\end{tabular}
\end{table*}


\begin{thebibliography}{52}
\bibitem[Acerbi et al.(1997)]{Ace97} Acerbi, F., Blättler, E., Dalmazio, D., Dedoch, A., Diethelm, R., Krobusek, B., Kohl, M., Locher, K., Martignoni, M., Pampaloni, C., Paschke, A., Peter, H., 1997, BBSAG, 114, 1
\bibitem[Agerer \& Hubscher(1995)]{Age95} Agerer, F., Hubscher, J., 1995, IBVS, 4222, 1
\bibitem[Azarnova(1956)]{Aza56} Azarnova, T.A., 1956, PZ, 11, 316
\bibitem[Baldwin \& Samolyk(1996)]{Bal96} Baldwin, M.E., Samolyk, G., 1996, "Observed Minima Timings of Eclipsing Binaries, No 3", Cambridge, MA (USA), American Association of Variable Star Observers. ISBN 1-878174-27-4
\bibitem[Baldwin(2007)]{Bal07} Baldwin, M.E., 2007, "Observed Minima Timings of Eclipsing Binaries, No 12", Cambridge, MA (USA), American Association of Variable Star Observers. ISBN 1-878174-75-4
\bibitem[Bl\"{a}ttler et al.(1996)]{Bla96} Bl\"{a}ttler, E., Dalmazio, D., Dedoch, A., Diethelm, R., Locher, K., Martignoni, M., Paschke, A., Peter, H., 1996, BBSAG, 113, 1
\bibitem[Bl\"{a}ttler et al.(2000)]{Bla00} Bl\"{a}ttler, E., Diethelm, R., Locher, K., Paschke, A., Vandenbroere, J., Verrot, J., 2000, BBSAG, 121
\bibitem[Bl\"{a}ttler et al.(2001)]{Bla01} Bl\"{a}ttler, E., Diethelm, R., Guilbault, P., Locher, K., Paschke, A., Weilenmann, W., 2001, BBSAG, 126, 1.
\bibitem[Brancewicz \& Dworak(1980)]{Bra80} Brancewicz, H. K., Dworak, T. Z., 1980, AcA, 30, 501
\bibitem[Chun-Hwey et al.(2006)]{Chu06} Chun-Hwey, K., Chung-Uk, L., Yo-Na, Y., Sung-Soo, P., Duk-Hyon, K., Sang-Mok, C., Jang-Hee, W., 2006, IBVS, 5694, 1
\bibitem[Dawson \& Trapp(2004)]{Daw04} Dawson, B. \& Trapp, R. G., 2004, "In Basic and Clinical Biostatistics", The McGraw-Hill Companies Inc. Press, USA, p.61, p.134, p.245
\bibitem[Diethelm(2010)]{Die10} Diethelm, R., 2010, IBVS, 5920, 1
\bibitem[Diethelm(2011)]{Die11} Diethelm, R., 2011, IBVS, 5960, 1
\bibitem[Diethelm(2013)]{Die13} Diethelm, R., 2013, IBVS, 6042, 1
\bibitem[Girardi et al.(2000)]{Gir00} Girardi, L., Bressan, A., Bertelli, G., Chiosi, C., 2000, A\&AS 141, 371
\bibitem[Giuricin et al.(1983)]{Giu83} Giuricin, G., Mardirossian, F., Mezzetti, M., 1983, A\&AS, 54, 211
\bibitem[Green et al.(1999)]{Gre99} Green, S. B., Salkind, N. J., Akey, T. M., 1999, "Using SPSS for Windows: Analyzing and Understanding Data", Upper Saddle River, N.J., London Prentice Hall Press, P.50
\bibitem[Handler \& Shobbrook(2002)]{Han02} Handler, G., Shobbrook, R. R., 2002, MNRAS, 333, 251
\bibitem[Henry et al.(2005)]{Hen05} Henry, G. W., Fekel, F. C., Henry, S. M., 2005, AJ, 129, 2815
\bibitem[Hubscher et al.(2006)]{Hub06} Hubscher, J., Paschke, A., Walter, F., 2006, IBVS, 5731, 1
\bibitem[Hubscher \& Walter(2007)]{Hub07} Hubscher, J., Walter, F., 2007, IBVS, 5761, 1
\bibitem[Hubscher et al.(2009a)]{Hub09a} Hubscher, J., Steinbach, H., Walter, F., 2009a, IBVS, 5874, 1
\bibitem[Hubscher et al.(2009b)]{Hub09b} Hubscher, J., Steinbach, H., Walter, F., 2009b, IBVS, 5889, 1
\bibitem[Hubscher et al.(2010a)]{Hub10a} Hubscher, J., Lehmann, P.B., Monninger, G., Steinbach, H., Walter, F., 2010a, IBVS, 5918, 1
\bibitem[Hubscher et al.(2010b)]{Hub10b} Hubscher, J., Lehmann, P.B., Monninger, G., Steinbach, H., Walter, F., 2010b, IBVS, 5941, 1
\bibitem[Hubscher(2011)]{Hub11} Hubscher, J, 2011, IBVS, 5984, 1
\bibitem[Hubscher et al.(2012a)]{Hub12a} Hubscher, J., Lehmann, P.B., Walter, F., 2012a, IBVS, 6010, 1
\bibitem[Hubscher \& Lehmann(2012b)]{Hub12b} Hubscher, J., Lehmann, P.B., 2012b, IBVS, 6026, 1
\bibitem[\.{I}bano\v{g}lu et al.(2007)]{Iba07} \.{I}bano\v{g}lu, C., Ta\c{s}, G., Sipahi, E., Evren, S., 2007, MNRAS, 376, 573
\bibitem[\.{I}bano\v{g}lu et al.(2008)]{Iba08} \.{I}bano\v{g}lu, C., Evren, S., Ta\c{s}, G., \c{C}ak{\i}rl{\i}, \"{O}., Bozkurt, Z., Af\c{s}ar, M., Frasca, A., Sipahi, E., Dal, H. A., \"{O}zdarcan, O., \c{C}amurdan, D. Z., \c{C}amurdan, M., 2008, MNRAS, 384, 331
\bibitem[Kaye et al.(1999)]{Kay99} Kaye, A. B., Handler, G., Krisciunas, K., Poretti, E., Zerbi, E., 1999, PASP, 111, 840
\bibitem[Kim et al.(2006)]{Kim06} Kim, C. H., Lee, C. U., Yoon, Y. N., Park, S. S., Kim, D. H., Cha, S. M., Won, J. H., 2006, IBVS, 5694, 1
\bibitem[Kjurkchieva \& Ivanov(2006)]{Kju06} Kjurkchieva, D.P., Ivanov, V.P., 2006, Bulgarian Astronomical Journal, Vol. 8, p. 57
\bibitem[Kopal(1978)]{Kop78} Kopal, Z., 1978, "Dynamics of Close Binary Systems", D. Reidel Pup. Com., Dordrecht: Holland/Boston, USA. London: England.
\bibitem[Kreiner(2004)]{Kre04} Kreiner, J. M., 2004, AcA, 54, 207
\bibitem[Lenz \& Breger(2005)]{Len05} Lenz, P., Breger, M., 2005, Comm. Asteroseismology, 146, 53
\bibitem[Lucy(1967)]{Luc67} Lucy, L. B., 1967, Z. Astrophys, 65, 89
\bibitem[Malkov et al.(2006)]{Mal06} Malkov, O. Y., Oblak, E., Snegireva, E. A., Torra, J., 2006, A\&A, 446, 785
\bibitem[Motulsky(2007)]{Mot07} Motulsky, H., 2007, "In GraphPad Prism 5: Statistics Guide", GraphPad Software Inc. Press, San Diego CA, p.94, p.133
\bibitem[Paschke \& Br\'{a}t(2006)]{Pas06} Paschke, A., Br\'{a}t, L., 2006, OEJV, 23, 13
\bibitem[Pols et al.(1998)]{Pol98} Pols, O. R., Schroder K. P., Hurley, J. R., Tout, C. A., Eggleton, P., 1998, MNRAS, 298, 525
\bibitem[Pr\v{s}a \& Zwitter(2005)]{Pra05} Pr\v{s}a, A., Zwitter, T., 2005, ApJ, 628, 426
\bibitem[Rolland et al.(2002)]{Rol02} Rolland, A., Costa, V., Rodriguez, E., Amado, P.J., Garcia-Pelayo, J.M., Lopez de Coca, P., Olivares, I., 2002, CoAst, 142, 57
\bibitem[Rucinski(1969)]{Ruc69} Rucinski, S. M., 1969, AcA, 19, 245
\bibitem[Shaw(1994)]{Sha94} Shaw, J. S., 1994, MmSAI, 65, 95
\bibitem[Sipahi et al.(2009)]{Sip09} Sipahi, E., Dal, H. A., \"{O}zdarcan, O., 2009, IBVS, 5904, 1
\bibitem[Soydugan et al.(2006)]{Soy06} Soydugan, E., Soydugan, F., Demircan, O., \.{I}bano\v{g}lu, C., 2006, MNRAS, 370, 2013
\bibitem[Tokunaga(2000)]{Tok00} Tokunaga, A.T., 2000, "Allen's Astrophysical Quantities", Fouth Edition, ed. A.N.Cox (Springer), p.143
\bibitem[Wall \& Jenkins(2003)]{Wal03} Wall, J. W. \& Jenkins, C. R., 2003, "In Practical Statistics For Astronomers", Cambridge University Press, p.79
\bibitem[Wilson \& Devinney(1971)]{Wil71} Wilson, R. E., Devinney, E. J., 1971, ApJ, 166, 605
\bibitem[Wilson(1990)]{Wil90} Wilson, R. E., 1990, ApJ, 356, 613
\bibitem[Zacharias et al.(2004)]{Zac04} Zacharias, N., Monet, D. G., Levine, S. E., Urban, S. E., Gaume, R., Wycoff, G. L., 2004, AAS, 205, 4815
\end{thebibliography}
\end{document}